\documentclass[conference]{IEEEtran}

\IEEEsettopmargin{t}{0.755in}

\usepackage{cite}
\usepackage{graphicx}
\usepackage{subcaption}
\usepackage{xcolor}
\usepackage{pgfplots}
\usepackage{multirow}
\usepackage{upgreek}
\usepackage{amssymb}
\usepackage{amsmath, mathtools}
\usepackage{cases}
\usepackage{pifont}
\usepackage{bm}
\usepackage{amsthm}

\usepackage{tabularx}
\usepackage{booktabs}
\newcolumntype{Y}{>{\centering\arraybackslash}X}
\usepackage{array}
\usepackage{float}
\usepackage{xcolor}
\usepackage{algorithmic}
\usepackage[linesnumbered,ruled,vlined]{algorithm2e}
\usepackage{changes}
\usepackage{bbm}
\usepackage{soul}
\usepackage{physics}
\usetikzlibrary{arrows,positioning,shapes.geometric,spy}

\newcommand{\fixme}[2]{\ifx&#2&{\leavevmode\color{red}#1}\else{\leavevmode\color{red}FIXME\{}#1{\leavevmode\color{red}\}}\footnote{{\leavevmode\color{red}#2}}\PackageWarning{Fixme}{#1: #2}\fi}

\SetCommentSty{mycommfont}

\DeclareMathOperator*{\argmax}{arg\,max}
\DeclareMathOperator*{\sgn}{sgn}

\DeclareMathOperator*{\CABP}{CABP}
\DeclareMathOperator*{\Beta}{Beta}
\DeclareMathOperator*{\VerifyCRC}{VerifyCRC}
\DeclareMathOperator*{\Algo}{Algo}
\DeclareMathOperator*{\InitBandit}{InitBandit}
\DeclareMathOperator*{\SelectAction}{SelectAction}
\DeclareMathOperator*{\UpdateBandit}{UpdateBandit}

\begin{document}

\title{Decoding Polar Codes with Reinforcement Learning}

\author{
\IEEEauthorblockN{Nghia Doan\IEEEauthorrefmark{1}, Seyyed Ali Hashemi\IEEEauthorrefmark{2}, Warren J. Gross\IEEEauthorrefmark{1}}
\IEEEauthorblockA{\IEEEauthorrefmark{1}Department of Electrical and Computer Engineering, McGill University, Canada} 
\IEEEauthorblockA{\IEEEauthorrefmark{2}Department of Electrical Engineering, Stanford University, USA}
\IEEEauthorblockA{nghia.doan@mail.mcgill.ca, ahashemi@stanford.edu, warren.gross@mcgill.ca}
}

\maketitle
\begin{abstract}

In this paper we address the problem of selecting factor-graph permutations of polar codes under belief propagation (BP) decoding to significantly improve the error-correction performance of the code. In particular, we formalize the factor-graph permutation selection as the multi-armed bandit problem in reinforcement learning and propose a decoder that acts like an online-learning agent that learns to select the good factor-graph permutations during the course of decoding. We use state-of-the-art algorithms for the multi-armed bandit problem and show that for a 5G polar codes of length $128$ with $64$ information bits, the proposed decoder has an error-correction performance gain of around $0.125$ dB at the target frame error rate of $10^{-4}$, when compared to the approach that randomly selects the factor-graph permutations.

\end{abstract}

\IEEEpeerreviewmaketitle
\section{Introduction}
\label{sec:intro}

Polar codes are a breakthrough in the field of channel coding as they can achieve the capacity of any binary symmetric channel with efficient encoding and decoding algorithms \cite{arikan}. Successive cancellation (SC) and belief propagation (BP) decoding algorithms were introduced in \cite{arikan} to decode polar codes. Although SC decoding can provide a low-complexity implementation, its serial nature prevents the decoder to reach a high decoding throughput. Furthermore, the error-correction performance of SC decoding for short to moderate polar codes does not satisfy the requirements of the fifth generation of cellular mobile communications (5G) standard. To improve the error-correction performance of SC decoding, SC list (SCL) decoding was introduced in \cite{tal_list} and it was shown that SCL can provide a significant error-correction performance improvement if it is concatenated with a cyclic redundancy check (CRC). Based on this observation, polar codes have been selected to be used in the enhanced mobile broadband (eMBB) control channel of 5G together with a CRC \cite{3gpp_report}.

Unlike SC-based decoders, the iterative message passing process of BP decoding can be executed in parallel, hence enabling the decoder to reach high decoding throughput. However, the conventional BP decoding algorithm suffers from poor error-correction performance. It has been shown that if polar codes are concatenated with a CRC, the error-correction performance of them under BP decoding can be significantly improved by exploiting the extrinsic information between the factor graphs of polar codes and the CRC \cite{Doan_ICC19, CABPList}. In addition, by using multiple independent permutations of the factor-graph of polar codes, the error-correction performance of them under BP decoding is significantly improved \cite{hussami2009performance, elkelesh2018belief, Doan_GLOBECOM, CABPList, LoopSimp}. However, the selection of good factor-graph permutations for polar codes that result in a correctly decoded codeword given a specific channel output realization remains an open research problem.

In this paper, we first formalize the selection of factor-graph permutations of polar codes under the CRC-aided (CA) BP (CABP) decoder in \cite{Doan_ICC19} as a multi-armed bandit problem in reinforcement learning (RL). We then utilize state-of-the-art algorithms designed for the multi-armed bandit problem to select the factor-graph permutations of polar codes that work best under CABP decoding. Unlike existing approaches, such as using genetic algorithm \cite{CABPList} or Monte Carlo-based methods \cite{Doan_GLOBECOM, LoopSimp}, in which the mechanism for the selection of factor-graph permutations requires off-line training, the proposed approach treats the CABP-based decoding of polar codes as an online-learning agent that learns to select the good factor-graph permutations during the course of decoding. We show that for a 5G polar code of length $128$ with $64$ information bits and concatenated with a $16$-bit 5G CRC, the proposed RL-aided CABP (RL-CABP) decoding algorithm has an error-correction performance gain of around $0.125$ dB, at the target frame error rate (FER) of $10^{-4}$, compared to the approach that selects the factor-graph permutations of polar codes randomly.

The remainder of the paper is as follows. Section~\ref{sec:polar} provides background on polar codes and BP-based decoding algorithms. Section~\ref{sec:bandit} summarizes the multi-armed bandit problem and its state-of-the-art algorithms. Section~\ref{sec:RL-CABP} introduces the proposed decoding algorithm, followed by the experimental results provided in Section~\ref{sec:exp}. Finally, concluding remarks are drawn in Section~\ref{sec:conclude}.

\section{Polar Codes}
\label{sec:polar}

\subsection{Polar Encoding}
A polar code $\mathcal{P}(N,K)$ of length $N$ with $K$ information bits is constructed by applying a linear transformation to the binary message word $\bm{u} = \{u_0,u_1,\ldots,u_{N-1}\}$ as $\bm{x} = \bm{u}\bm{G}^{\otimes n}$ where $\bm{x} = \{x_0,x_1,\ldots,x_{N-1}\}$ is the codeword, $\bm{G}^{\otimes n}$ is the $n$-th Kronecker power of the polarizing matrix $\bm{G}=\bigl[\begin{smallmatrix} 1&0\\ 1&1 \end{smallmatrix} \bigr]$, and $n = \log_2 N$. The vector $\bm{u}$ contains a set $\mathcal{I}$ of $K$ information bit indices and a set $\mathcal{I}^c$ of $N-K$ frozen bit indices. The positions of the frozen bits are known to the encoder and the decoder and their values are set to $0$. The codeword $\bm{x}$ is then modulated and sent through the channel. In this paper, binary phase-shift keying (BPSK) modulation and additive white Gaussian noise (AWGN) channel model are considered. Therefore, the soft vector of the transmitted codeword received by the decoder is written as ${\bm{y}=(\mathbf{1}-2\bm{x})+\bm{z}}$, where $\mathbf{1}$ is an all-one vector of size $N$, and $\bm{z} \in \mathbbm{R}^N$ is a Gaussian noise vector with variance $\sigma^2$ and zero mean. In the log-likelihood ratio (LLR) domain, the LLR vector of the transmitted codeword is given as
$
{\bm{L} = \ln{\frac{\text{Pr}(\bm{x}=0|\bm{y})}{\text{Pr}(\bm{x}=1|\bm{y})}}=\frac{2\bm{y}}{\sigma^2}}
$.

\subsection{Belief Propagation Decoding of Polar Codes}
\label{sec:polar:BPD}

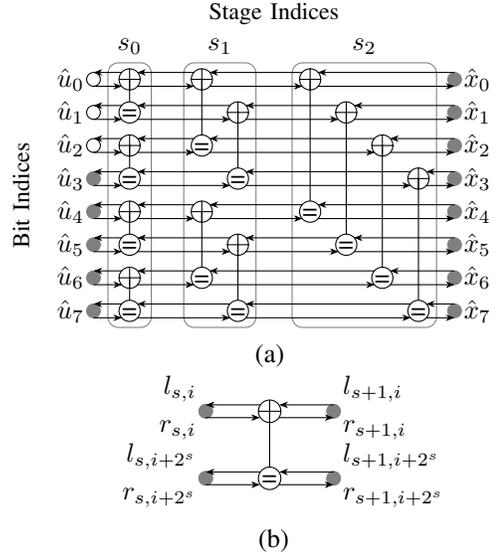
\begin{figure}[t]
	\centering
	\usetikzlibrary{arrows.meta}
\usetikzlibrary{positioning}

\centering
\begin{tikzpicture}[scale=0.8]
\def\xM{6}
\def\xs{\xM/10}
\def\ys{0.55}
\def\yo{0.115}
\def\xo{0.125}
\def\gain{1.075}
\def\markSize{3.25}

\node[] at (5*\xs, 9*\ys) {\small{Stage Indices}};
\node[rotate=90] at (-2*\xs, 3.5*\ys) {\small{Bit Indices}};


\draw[rounded corners, gray] (0.4*\xs,-0.5*\ys) rectangle (1.6*\xs,7.5*\ys);
\node[] at (1*\xs, 8*\ys) {$s_0$};

\draw[rounded corners, gray] (2.5*\xs, -0.5*\ys) rectangle (4.5*\xs, 7.5*\ys);
\node[] at (3.5*\xs, 8*\ys) {$s_1$};

\draw[rounded corners, gray] (5.5*\xs, -0.5*\ys) rectangle (9.5*\xs, 7.5*\ys);
\node[] at (7.5*\xs, 8*\ys) {$s_2$};

\foreach \i in{0,...,7}
{
	\pgfmathsetmacro\bIndex{int(7-\i)};
	
	\ifthenelse{\i < 5}{\draw[] plot[mark=*, mark size = \markSize, mark options={color=gray}] coordinates {(0*\xs,\i*\ys)}		
		node[left]{$\hat{u}_{\bIndex}$};}{\draw[] plot[mark=*, mark size = \markSize, mark options={fill=white}] coordinates {(0*\xs,\i*\ys)} node[left]{$\hat{u}_{\bIndex}$};};
	
	\draw[] plot[mark=*, mark size = \markSize, mark options={color=gray}] coordinates {(10*\xs,\i*\ys)} node[right]{$\hat{x}_{\bIndex}$};
}

\draw[] plot[mark=oplus, mark size = 5, mark options={fill=white}] coordinates {(\xs,\ys)} --
plot[mark=*, mark size = 5, mark options={fill=white}] coordinates {(\xs,0)} node[above=-0.225]{=};

\draw[] plot[mark=oplus, mark size = 5, mark options={fill=white}] coordinates {(\xs,3*\ys)} --
plot[mark=*, mark size = 5, mark options={fill=white}] coordinates {(\xs,2*\ys)} node[above=-0.225]{=};

\draw[] plot[mark=oplus, mark size = 5, mark options={fill=white}] coordinates {(\xs,5*\ys)} --
plot[mark=*, mark size = 5, mark options={fill=white}] coordinates {(\xs,4*\ys)} node[above=-0.225]{=};

\draw[] plot[mark=oplus, mark size = 5, mark options={fill=white}] coordinates {(\xs,7*\ys)} --
plot[mark=*, mark size = 5, mark options={fill=white}] coordinates {(\xs,6*\ys)} node[above=-0.225]{=};

\foreach \i in{0,...,7}
{
	\draw[->,>={Stealth[scale=0.7]}] (0*\xs, \i*\ys-\yo) -- (1*\xs-\xo, \i*\ys-\yo);
	\draw[<-,>={Stealth[scale=0.7]}] (0*\xs, \i*\ys+\yo) -- (1*\xs-\xo, \i*\ys+\yo);
}

\draw[] plot[mark=oplus, mark size = 5, mark options={fill=white}] coordinates {(3*\xs,3*\ys)} --
plot[mark=*, mark size = 5, mark options={fill=white}] coordinates {(3*\xs,\ys)} node[above=-0.225]{=};

\draw[] plot[mark=oplus, mark size = 5, mark options={fill=white}] coordinates {(3*\xs,7*\ys)} --
plot[mark=*, mark size = 5, mark options={fill=white}] coordinates {(3*\xs,5*\ys)} node[above=-0.225]{=};

\draw[] plot[mark=oplus, mark size = 5, mark options={fill=white}] coordinates {(4*\xs,2*\ys)} --
plot[mark=*, mark size = 5, mark options={fill=white}] coordinates {(4*\xs,0)} node[above=-0.225]{=};

\draw[] plot[mark=oplus, mark size = 5, mark options={fill=white}] coordinates {(4*\xs,6*\ys)} --
plot[mark=*, mark size = 5, mark options={fill=white}] coordinates {(4*\xs,4*\ys)} node[above=-0.225]{=};

\foreach \i in{1,3,5,7}
{
	\draw[->,>={Stealth[scale=0.7]}] (1*\xs+\xo, \i*\ys-\yo) -- (3*\xs-\xo, \i*\ys-\yo);
	\draw[<-,>={Stealth[scale=0.7]}] (1*\xs+\xo, \i*\ys+\yo) -- (3*\xs-\xo, \i*\ys+\yo);
}

\foreach \i in{0,2,4,6}
{
	\draw[->,>={Stealth[scale=0.7]}] (1*\xs+\xo, \i*\ys-\yo) -- (4*\xs-\xo, \i*\ys-\yo);
	\draw[<-,>={Stealth[scale=0.7]}] (1*\xs+\xo, \i*\ys+\yo) -- (4*\xs-\xo, \i*\ys+\yo);
}

\draw[] plot[mark=oplus, mark size = 5, mark options={fill=white}] coordinates {(9*\xs,4*\ys)} --
plot[mark=*, mark size = 5, mark options={fill=white}] coordinates {(9*\xs,0*\ys)} node[above=-0.225]{=};

\draw[] plot[mark=oplus, mark size = 5, mark options={fill=white}] coordinates {(8*\xs,5*\ys)} --
plot[mark=*, mark size = 5, mark options={fill=white}] coordinates {(8*\xs,1*\ys)} node[above=-0.225]{=};

\draw[] plot[mark=oplus, mark size = 5, mark options={fill=white}] coordinates {(7*\xs,6*\ys)} --
plot[mark=*, mark size = 5, mark options={fill=white}] coordinates {(7*\xs,2*\ys)} node[above=-0.225]{=};

\draw[] plot[mark=oplus, mark size = 5, mark options={fill=white}] coordinates {(6*\xs,7*\ys)} --
plot[mark=*, mark size = 5, mark options={fill=white}] coordinates {(6*\xs,3*\ys)} node[above=-0.225]{=};

\foreach \i in{0,4}
{
	\draw[->,>={Stealth[scale=0.7]}] (4*\xs+\xo, \i*\ys-\yo) -- (9*\xs-\xo, \i*\ys-\yo);
	\draw[<-,>={Stealth[scale=0.7]}] (4*\xs+\xo, \i*\ys+\yo) -- (9*\xs-\xo, \i*\ys+\yo);
}

\foreach \i in{1,5}
{
	\draw[->,>={Stealth[scale=0.7]}] (3*\xs+\xo, \i*\ys-\yo) -- (8*\xs-\xo, \i*\ys-\yo);
	\draw[<-,>={Stealth[scale=0.7]}] (3*\xs+\xo, \i*\ys+\yo) -- (8*\xs-\xo, \i*\ys+\yo);
}

\foreach \i in{2,6}
{
	\draw[->,>={Stealth[scale=0.7]}] (4*\xs+\xo, \i*\ys-\yo) -- (7*\xs-\xo, \i*\ys-\yo);
	\draw[<-,>={Stealth[scale=0.7]}] (4*\xs+\xo, \i*\ys+\yo) -- (7*\xs-\xo, \i*\ys+\yo);
}

\foreach \i in{3,7}
{
	\draw[->,>={Stealth[scale=0.7]}] (3*\xs+\xo, \i*\ys-\yo) -- (6*\xs-\xo, \i*\ys-\yo);
	\draw[<-,>={Stealth[scale=0.7]}] (3*\xs+\xo, \i*\ys+\yo) -- (6*\xs-\xo, \i*\ys+\yo);
}


\foreach \i in{0,4}
{
	\draw[->,>={Stealth[scale=0.7]}] (9*\xs+\xo, \i*\ys-\yo) -- (10*\xs, \i*\ys-\yo);
	\draw[<-,>={Stealth[scale=0.7]}] (9*\xs+\xo, \i*\ys+\yo) -- (10*\xs, \i*\ys+\yo);
}

\foreach \i in{1,5}
{
	\draw[->,>={Stealth[scale=0.7]}] (8*\xs+\xo, \i*\ys-\yo) -- (10*\xs, \i*\ys-\yo);
	\draw[<-,>={Stealth[scale=0.7]}] (8*\xs+\xo, \i*\ys+\yo) -- (10*\xs, \i*\ys+\yo);
}

\foreach \i in{2,6}
{
	\draw[->,>={Stealth[scale=0.7]}] (7*\xs+\xo, \i*\ys-\yo) -- (10*\xs, \i*\ys-\yo);
	\draw[<-,>={Stealth[scale=0.7]}] (7*\xs+\xo, \i*\ys+\yo) -- (10*\xs, \i*\ys+\yo);
}

\foreach \i in{3,7}
{
	\draw[->,>={Stealth[scale=0.7]}] (6*\xs+\xo, \i*\ys-\yo) -- (10*\xs, \i*\ys-\yo);
	\draw[<-,>={Stealth[scale=0.7]}] (6*\xs+\xo, \i*\ys+\yo) -- (10*\xs, \i*\ys+\yo);
}
\end{tikzpicture}\\
\hspace*{10pt} (a)

\hspace*{20pt}
\begin{tikzpicture}[scale=.85]

\usetikzlibrary{arrows.meta}
\usetikzlibrary{positioning}

\def\markSize{3}

\draw[] plot[mark=*, mark size = \markSize, mark options={color=gray}] coordinates {(0,0.75)} node[above left] {$l_{s,i}$};

\draw[] plot[mark=*, mark size = \markSize, mark options={color=gray}] coordinates {(0,0.75)} node[below left] {$r_{s,i}$};

\draw[]plot[mark=*, mark size = \markSize, mark options={color=gray}] coordinates {(2,0.75)} node[above right] {$l_{s+1,i}$};

\draw[]plot[mark=*, mark size = \markSize, mark options={color=gray}] coordinates {(2,0.75)} node[below right] {$r_{s+1,i}$};

\draw[] plot[mark=*, mark size = \markSize, mark options={color=gray}] coordinates {(0,-0.3)} node[above left] {$l_{s,i+2^s}$};

\draw[] plot[mark=*, mark size = \markSize, mark options={color=gray}] coordinates {(0,-0.3)} node[below left] {$r_{s,i+2^s}$};

\draw[] plot[mark=*, mark size = \markSize, mark options={color=gray}] coordinates {(2,-0.3)} node[above right] {$l_{s+1,i+2^s}$};

\draw[] plot[mark=*, mark size = \markSize, mark options={color=gray}] coordinates {(2,-0.3)} node[below right] {$r_{s+1,i+2^s}$};

\draw[] plot[mark=oplus, mark size = 5, mark options={fill=white}] coordinates {(1,0.75)} --
plot[mark=*, mark size = 5, mark options={fill=white}] coordinates {(1,-0.3)} node[above=-0.225]{=};

\draw[<-,>={Stealth[scale=0.7]}] (0,-0.2) -- (0.85,-0.2);
\draw[->,>={Stealth[scale=0.7]}] (0,-0.4) -- (0.85,-0.4);

\draw[<-,>={Stealth[scale=0.7]}] (1.15,-0.2) -- (2,-0.2);
\draw[->,>={Stealth[scale=0.7]}] (1.15,-0.4) -- (2,-0.4);

\draw[<-,>={Stealth[scale=0.7]}] (0,0.85) -- (0.85,0.85);
\draw[->,>={Stealth[scale=0.7]}] (0,0.65) -- (0.85,0.65);

\draw[<-,>={Stealth[scale=0.7]}] (1.15,0.85) -- (2,0.85);
\draw[->,>={Stealth[scale=0.7]}] (1.15,0.65) -- (2,0.65);

\end{tikzpicture}\\
\hspace*{20pt} (b)		
	\vspace*{5pt}
	\caption{(a) Factor-graph representation of $\mathcal{P}(8,5)$ with $\mathcal{I}^c = \{0,1,2\}$, (b) a PE for BP decoding.}
	\label{fig:BPGraph}
	\vspace*{-1\baselineskip}
\end{figure}

Fig.~\ref{fig:BPGraph}a illustrates BP decoding on a factor graph representation of $\mathcal{P}(8,5)$. The messages are iteratively propagated through the processing elements (PEs) \cite{arikan2010polar}. An update iteration starts with a right-to-left message pass that propagates the LLR values from the channel stage (right-most stage), to the information bit stage (left-most stage), and ends with the left-to-right message pass occurring in the reverse order. Fig.~\ref{fig:BPGraph}b shows a PE with its corresponding messages, where $r_{s,i}$ denotes a left-to-right message, and $l_{s,i}$ denotes a right-to-left message of the $i$-th bit index at stage $s$. The update rule for the right-to-left messages of a PE is \cite{arikan2010polar}
\begin{align}
\label{PolarPE_left}
\begin{split}
\begin{cases}
l_{s,i} &= f(l_{s+1,i},r_{s,i+2^s} + l_{s,i+2^s})\text{,}\\
l_{s,i+2^s} &= f(l_{s+1,i},r_{s,i}) + l_{s+1,i+2^s}\text{,}\\
\end{cases}
\end{split}
\end{align}
and for the left-to-right messages is
\begin{align}
\label{PolarPE_right}
\begin{split}
\begin{cases}
r_{s+1,i} &= f(r_{s,i},l_{s+1,i+2^s} + r_{s,i+2^s})\text{,}\\
r_{s+1,i+2^s} &= f(r_{s,i},l_{s+1,i}) + r_{s,i+2^s}\text{,}
\end{cases}
\end{split}
\end{align}
where $f(.)$ is the scaled min-sum function \cite{yuan2014early}:
\begin{equation}
\label{minsum}
f(x,y) = 0.9375\times\sgn(x)\sgn(y)\min(|x|,|y|)\text{.}
\end{equation}

The BP decoding performs a predetermined $I_{\max}$ iterations where the messages are propagated through all PEs in accordance with (\ref{PolarPE_left}) and (\ref{PolarPE_right}). The LLR values at stage $0$, denoted as $\bm{r}_0$, are initialized as
\begin{equation}
r_{0,i} =
\begin{cases}
0 \text{,} & \text{if } i \in \mathcal{I} \text{,}\\
+\infty \text{,} & \text{if } i \in \mathcal{I}^c \text{,}
\end{cases}
\end{equation}
and the LLR values at stage $n$, denoted as $\bm{l}_n$, are initialized as $\bm{l}_n=\bm{L}$. In addition, all the other left-to-right and right-to-left messages of the PEs at the first iteration are set to $0$. After running $I_\text{max}$ iterations, the decoder makes a hard decision on the LLR values of the $i$-th bit at the information bit stage to obtain the estimated message word as
\begin{equation}
\label{hardDec}
\hat{u}_i=
\begin{cases}
0 \text{,} & \text{if } r_{0,i} + l_{0,i} \geq 0 \text{,}\\
1 \text{,} & \text{otherwise.}
\end{cases} 
\end{equation}	

In this paper we consider the case where a CRC is concatenated to the polar code as in the 5G standard. After running BP decoding on the factor-graph of polar codes for $I_\text{min}$ iterations ${(0 < I_\text{min} < I_\text{max})}$, a CRC verification is performed to early-terminate the decoding process. In addition, the factor-graph of CRC is utilized to further improve the error-correction performance of polar codes under BP decoding in a way that the extrinsic information of the factor-graphs of CRC and polar codes is exchanged by running BP decoding on both factor-graphs after the $I_\text{min}$-th iteration \cite{Doan_ICC19}. We refer to this algorithm as CABP decoding.

\subsection{Decoding Polar Codes on Factor-Graph Permutations}

The error-correction performance of polar codes under different decoding algorithms can significantly improve if the decoding is performed independently on multiple factor-graph permutations \cite{hussami2009performance, elkelesh2018belief, Doan_GLOBECOM, CABPList, LoopSimp}. A factor-graph permutation, denoted as $\pi_p$ $(0 \leq p < n! )$, is constructed by permuting the PE stages of the polar codes factor graph \cite{hussami2009performance}. For instance, Fig.~\ref{fig:BPGraph}a shows the original factor graph of $\mathcal{P}(8,5)$, denoted as $\pi_0=\{s_0,s_1,s_2\}$. Permuting the PEs in stage $s_1$ and $s_2$ in Fig.~\ref{fig:BPGraph}a forms another factor-graph permutation, $\pi_1=\{s_0,s_2,s_1\}$. It was shown that there is a one-to-one mapping between the factor-graph permutation and the bit-index permutation of the original factor-graph \cite{Doan_GLOBECOM}. Thus, the decoding of polar codes on their permuted factor graphs can be performed by running the decoder on the permuted bit-indices of the original factor graph. This keeps the architecture of the decoder unchanged \cite{Doan_GLOBECOM}.

In this paper, given  $\pi_p$ and $\bm{L}$, we use the technique presented in \cite{Doan_GLOBECOM} to form the corresponding permuted bit-indices of the channel LLR values, $\bm{L}_{\pi_p}$. We then apply CABP decoding on $\bm{L}_{\pi_p}$ using the original factor-graph. Note that the permuted soft messages of the information bit stage $\bm{l}_{0_{\pi_p}}$ is permuted back to $\bm{l}_0$ before running BP decoding on the CRC factor-graph. Given $\bm{L}$ and $\pi_p$, we consider CABP decoding as a function and denote its output as $\bm{\hat{u}}=\CABP(\bm{L}, \pi_p)$. In addition, throughout this paper, we refer to $\pi_0$ as the permutation corresponding to the original factor-graph.

\section{Multi-Armed Bandit Problem}
\label{sec:bandit}

A multi-armed bandit problem, or a $k$-armed bandit problem $(k>1)$, is an RL problem where an agent has to repeatedly make a choice among $k$ different actions (options). After each action is performed, the agent receives a numerical reward that is drawn from a distribution that depends on the selected action. The agent's objective is to maximize the expected cumulative rewards over a time period \cite{Sutton}. Let ${\mathcal{A}=\{a_1,a_2,\ldots,a_k\}}$ be the set of actions and $q^*(a_j)$ $(1\leq j \leq k)$ be the corresponding expected reward of an action $a_j$. $q^*(a_j)$ is called the value function and its value is unknown to the agent. In this paper, we consider three state-of-the-art algorithms designed for the multi-armed bandit problem, namely, $\varepsilon$-greedy, upper confidence bound (UCB), and Thompson sampling (TS).

\subsection{$\varepsilon$-Greedy and UCB Algorithms}

Let $n_{a_j}$ be the number of times that an action $a_j$ is selected up to the $t$-th time step. If $a_j$ is selected at the $t$-th time step, $n_{a_j}$ is updated as ${n_{a_j}:=n_{a_j}+1}$ \cite{Sutton}. Then, the value function $q^*(a_j)$ is estimated as $Q_{a_j}$ in accordance with ${Q_{a_j} := Q_{a_j} + \frac{1}{n_{a_j}}\left[R_t-Q_{a_j}\right]}$, where $R_t$ is the reward received by selecting action $a_j$ at the $t$-th time step \cite{Sutton}. Initially, $Q_{a_j}$ and $a_j$ are set to 0 $(\forall j, 1 \leq j \leq k)$. Given the estimated expected rewards $Q_{a_j}$, an exploitation occurs when the agent selects an action that has the largest expected reward value \cite{Sutton}. On the other hand, an exploration occurs when the agent selects any action that does not have the largest expected reward value \cite{Sutton}. 

Let $a_{j^*}$ be the action selected by the agent at the $t$-th time step. Under the $\varepsilon$-greedy algorithm $a_{j^*}$ is selected as \cite{Sutton}
\begin{equation}
a_{j^*} =
\begin{cases}
\displaystyle\argmax_{\forall {a_j}} Q_{a_j} &\text{with probability $1-\varepsilon$,}\\
a_\text{random} &\text{with probability $\varepsilon$,}\\
\end{cases}
\end{equation}
where $a_\text{random}$ is a random action drawn i.i.d. from $\mathcal{A}$. On the other hand, under the UCB algorithm $a_{j^*}$ is selected as
\begin{equation}
a_{j^*}=\argmax_{\forall a_j} \left[Q_{a_j}+c\sqrt{\frac{\ln t}{n_{a_j}}}\right],
\end{equation}
where $n_{a_j}\neq0$ and $c \in \mathbb{R}^+$. If $n_{a_j}=0$, $a_j$ is considered as an exploitation action. Note that $\varepsilon$ and $c$ control the degree of exploration of the $\varepsilon$-greedy and UCB algorithms, respectively.

\subsection{Thompson Sampling}

Instead of estimating the expected reward value $q^*(a_j)$ as in the $\varepsilon$-greedy and UCB algorithms, the TS algorithm directly estimates the distribution of the reward value associated with each action. In this paper, as $R_t\in\{0,1\}$ a Beta distribution is used to estimate the reward's distribution \cite{TS}. A Beta distribution has two shape parameters: $\alpha,\beta \in \mathbb{R}^+$, and a different set of shape parameters is used for each action. We denote a random sampling from the estimated reward distribution of the $j$-th action as $\upsilon_{a_j}=\Beta(\alpha_{a_j}, \beta_{a_j})$. At the $t$-th time step, the TS algorithm first draws a random sample from each of the estimated reward distributions. The agent then selects the action $a_{j^*}$ as $a_{j^*} = \argmax_{\forall {a_j}} \upsilon_{a_j}$. The shape parameters corresponding to the selected action $a_{j^*}$ are then updated as $\alpha_{a_{j^*}}:=\alpha_{a_{j^*}}+R_t$ and $\beta_{a_{j^*}}:=\beta_{a_{j^*}}+R_t$ \cite{TS}. Initially, ${\alpha_{a_j}=\beta_{a_j}=1}$ ${(\forall j, 1 \leq j \leq k)}$ \cite{TS}.

\section{Selection of Factor-Graph Permutations with Reinforcement Learning}
\label{sec:RL-CABP}

This section first formalizes the selection of factor-graph permutations for polar decoding as a $k$-armed bandit problem. It then introduces the proposed decoding method that utilizes the multi-armed bandit algorithms in Section~\ref{sec:bandit} to select the factor-graph permutations under CABP decoding.

\subsection{Problem Formulation}

Under BP decoding of polar codes, the  original factor-graph permutation $\pi_0$ is empirically observed to have the best error-correction performance compared to other factor-graph permutations \cite{hussami2009performance}. However, there are cases that a specific channel output realization, which cannot be decoded using the original factor-graph permutation, can be decoded using another factor-graph permutation \cite{hussami2009performance}. As the number of permutations, $n!$, is large, running BP decoding on all of the permutations is not possible in real applications. Instead, the decoding is performed on a small set of $M$ factor-graph permutations, including the original factor-graph permutation \cite{hussami2009performance,elkelesh2018belief,Doan_GLOBECOM,CABPList}.

Let an action $a_j \in \mathcal{A}$ $(1\leq j \leq k)$ be a random selection of $M-1$ $(M>1)$ factor-graph permutations that do not include the original factor-graph permutation. Consider the CRC verification is not successful when CABP decoding is performed on the original factor-graph permutation $\pi_0$. The proposed decoder then selects an action $a_j$ from the set $\mathcal{A}$. If one of the factor-graph permutations in $a_j$ results in a successful CRC verification, a reward of $1$ is given to the decoder. Otherwise, if none of the permutations in $a_j$ results in a successful CRC verification under CABP decoding, a reward of $0$ is given to the decoder. Therefore, among $k$ sets of predefined factor-graph permutations, i.e. $k$ different actions, the proposed decoding algorithm decides which set of factor-graph permutations maximizes the reward during the course of decoding. The selection of factor-graph permutations for CABP decoding can thus be formalized as a $k$-armed bandit problem as defined in Section~\ref{sec:bandit}.

\subsection{Reinforcement Learning-Aided CABP}
\label{sec:RL-CABPB}

\begin{algorithm}[t]
	\DontPrintSemicolon
	\caption{Forming the action set}
	\label{alg1}
	\SetKwInOut{Input}{Input}
	\SetKwInOut{Output}{Output}
	
	\Input{$n,k,M$}
	\Output{$\mathcal{A}$}
	
	\tcp{Define the original permutation}
	$\pi_0\leftarrow\{s_0,s_1,\cdots,s_{n-1}\}$\\
	\tcp{Select $M-1$ random permutations for each action}
	$\mathcal{A} \leftarrow \emptyset$\\
	\For{$j \leftarrow 1$ \KwTo $k$}
	{
		$a_j \leftarrow \emptyset$\\
		\For{$t \leftarrow 1$ \KwTo $M-1$}
		{
			$\pi_{j,t} \leftarrow \text{RandShuffle}(\pi_0)$\\
			$a_j \leftarrow a_j \cup \pi_{j,t}$\\
		}
		$\mathcal{A} \leftarrow \mathcal{A} \cup a_j$
	}
	\Return $\mathcal{A}$
\end{algorithm}

The proposed decoding algorithm starts with the construction of $\mathcal{A}$, the set of $k$ different actions, which is outlined in Algorithm~\ref{alg1}. Each action $a_j \in \mathcal{A}$ contains $M-1$ random factor-graph permutations. Formally, $a_j = \{\pi_{j,1},\pi_{j,2},\cdots,\pi_{j,M-1}\}$, ${\pi_{j,t} \neq \pi_0}$ $\forall j,t$,  where $1\leq j \leq k$, and $1 \leq t \leq M-1$. A random factor-graph permutation is formed by randomly permuting the PE stages of the original factor graph $\pi_0$, which is obtained by the $\text{RandShuffle}$ function in Algorithm~\ref{alg1}. The number of all possible actions is
\begin{equation}
\label{equ:k_max}
k_{\max}={{n!-1}\choose{M-1}}=\frac{(n!-1)!}{(M-1)!(n!-M)!},
\end{equation}
which is generally intractable for practical values of $n$ and $M$. Therefore, only the subset $\mathcal{A}$ of all the possible actions is considered. In fact, $\mathcal{A}$ is constructed by randomly sampling from the complete set of actions as shown in Algorithm~\ref{alg1}. Note that after $\mathcal{A}$ is formed, the set of actions in $\mathcal{A}$ remains unchanged during the course of decoding.

Algorithm~\ref{alg2} outlines the proposed RL-CABP decoding algorithm, given the predefined set of actions $\mathcal{A}$ constructed in Algorithm~\ref{alg1}. The proposed RL-CABP decoder first initializes the parameters of the multi-armed bandit algorithm depending on its type, which is defined by the parameter $\Algo$ in Algorithm~\ref{alg2}. If $\Algo$ indicates the $\varepsilon$-greedy or UCB algorithms, the parameters of the multi-armed bandit algorithm are initialized as $Q_{a_j}=n_{a_j}=0$ $\forall j$, $1 \leq j \leq k$. If the TS algorithm is used, the set of parameters is initialized as $\alpha_{a_j}=\beta_{a_j}=1$ $\forall j$, $1 \leq j \leq k$. Note that the initialization process is only carried out once in the course of decoding.

Then, the proposed RL-CABP decoding applies CABP decoding over the original factor-graph permutation $\pi_0$. If the CRC verification, which is obtained by the $\VerifyCRC$ function in Algorithm~\ref{alg2} is successful, the proposed decoder outputs the estimated message word $\bm{\hat{u}}$ and the decoding process is terminated. Otherwise, the RL-CABP decoder selects an action $a_{j^*}$ from $\mathcal{A}$, which contains a set of $M-1$ random factor-graph permutations as described in Algorithm~\ref{alg1}. Depending on the type of the algorithm, the function $\SelectAction$ implements the selection criteria of the considered multi-armed bandit algorithms as introduced in Section~\ref{sec:bandit}. Note that the $\SelectAction$ function can be performed in parallel with the first CABP decoding attempt as there is no dependency between them. Therefore, the selected action $a_{j^*}$ can be obtained in advance without adding a latency overhead to the proposed decoding algorithm. Moreover, if the first CABP decoding attempt over $\pi_0$ is successful, the selected action $a_{j^*}$ is discarded.

If the first CABP decoding attempt fails in the proposed RL-CABP decoding algorithm, additional CABP decoding attempts are sequentially carried over the factor-graph permutations specified by $a_{j^*}$. As soon as the CRC verification is successful after CABP decoding on one of the factor-graph permutations in $a_{j^*}$, a reward of $1$ is given to $a_{j^*}$, and the decoding outputs the estimated message word that satisfies the CRC verification. On the other hand, if running CABP on all of the permutations in $a_{j^*}$ does not result in a successful CRC test, a reward of $0$ is given to $a_{j^*}$ and the decoding is declared unsuccessful. Finally, after each action selection, the parameters associated with the selected action $a_{j^*}$ are updated using the $\UpdateBandit$ function. Note that the parameter update process is based on the received reward and the type of the mutli-armed bandit algorithm as provided in Section~\ref{sec:bandit}.

\begin{algorithm}[t]
	\DontPrintSemicolon
	\caption{RL-CABP Decoding}
	\label{alg2}
	\SetKwInOut{Input}{Input}
	\SetKwInOut{Output}{Output}
	
	\Input{$\bm{L}, \mathcal{A}, k, M, \Algo$}
	\Output{$\bm{\hat{u}}$}

	\tcp{Initialize the bandit parameters}
	$\InitBandit(k,\Algo)$\\
	
	\tcp{Apply CABP decoding on $\pi_0$}
	$\bm{\hat{u}} \leftarrow \CABP(\bm{L},\pi_0)$\\
    $isCorrect_{\pi_0} \leftarrow \VerifyCRC(\bm{\hat{u}})$\\

	\tcp{Select an action in advance}	
	$a_{j^*} \leftarrow \SelectAction (\mathcal{A},\text{Algo})$\\

	\tcp{If applicable, apply CABP decoding on the permutations specified by $a_{j^*}$}
	\If{$(isCorrect_{\pi_0} = 0)$}{
        $isCorrect_{a_{j^*}} \leftarrow 0$\\
		\For{$t \leftarrow 1$ \KwTo $M-1$}{
			$\bm{\hat{u}} \leftarrow \CABP(\bm{L},\pi_{j^*,t})$\\
			$isCorrect_{a_{j^*}} \leftarrow \VerifyCRC(\bm{\hat{u}})$\\
			\If{$(isCorrect_{a_{j^*}} = 1)$}{			
				\textbf{break}\\
			}
		}
		\tcp{Update the bandit parameters associated with $a_{j^*}$}
		$R_t \leftarrow isCorrect_{a_{j^*}}$\\
		$\UpdateBandit(R_t,a_{j^*},\Algo)$
	}
	\Return $\bm{\hat{u}}$
\end{algorithm}

\section{Experimental Results}
\label{sec:exp}

In this section, the performance of various multi-armed bandit algorithms used by the proposed RL-CABP decoding is numerically evaluated. In addition, the error-correction performance of the proposed RL-CABP decoding in terms of FER is compared with that of other polar decoding techniques. A complexity comparison of different multi-armed bandit algorithms in the proposed RL-CABP decoding is also given. We use $\mathcal{P}(128,64)$ selected for the eMBB control channel of the 5G standard \cite{3gpp_report}. Furthermore, the polar code is concatenated with a CRC of length $16$, which is also used in 5G \cite{3gpp_report}. The total number of factor-graph permutations used by all BP-based decoders is set to $7$. We set $I_{\max}=100$ and $I_{\min}=50$ for all BP-based decoding algorithms.

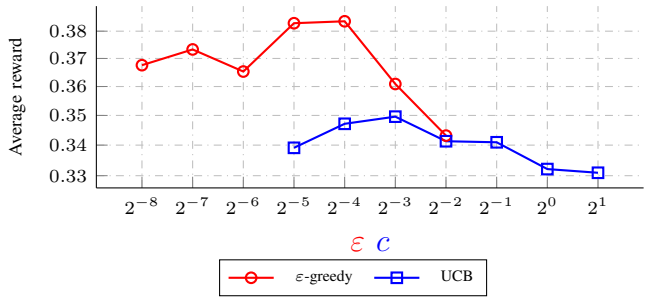
\begin{figure}[t!]
	\centering
	\begin{tikzpicture}
\pgfplotsset{
	label style = {font=\fontsize{7pt}{7}\selectfont},
	tick label style = {font=\fontsize{7pt}{7}\selectfont},
	log x ticks with fixed point/.style={
		xticklabel={
			\pgfkeys{/pgf/fpu=true}
			\pgfmathparse{exp(\tick)}%
			\pgfmathprintnumber[fixed relative, precision=3]{\pgfmathresult}
			\pgfkeys{/pgf/fpu=false}
		}
	},
	log y ticks with fixed point/.style={
		yticklabel={
			\pgfkeys{/pgf/fpu=true}
			\pgfmathparse{exp(\tick)}%
			\pgfmathprintnumber[fixed relative, precision=3]{\pgfmathresult}
			\pgfkeys{/pgf/fpu=false}
		}
	}
}

\begin{axis}[
ymode=log,
xmode=log,
log basis x=2,
yticklabel pos=left,
axis y line*=left,
ylabel style={yshift=-0.5em},
axis x line*=bottom,
ylabel=Average reward,
xlabel=\large{\color{red}{$\varepsilon$} \color{blue}{$c$}},
ytick={0.33,0.34,0.35,0.36,0.37,0.38,0.39,0.4},
xtick={3.90625E-3, 7.8125E-3,0.015625,0.03125,0.0625,0.125,0.25,0.5,1,2,4},
ymajorgrids=true,
xmajorgrids=true,
log y ticks with fixed point,
ylabel style = {align=center},
grid style=dashdotted,
width=1\columnwidth, height=4cm,
legend style={
	at={(0,1e-5)},
	anchor=south west,
	column sep= 2mm,
	font=\fontsize{6pt}{7.2}\selectfont,
},
legend to name=legend_param_study,
legend columns=2,
],

\addplot[
color=red,
solid,
mark=o,
thick,
]
table {
3.90625E-3 0.367614
7.8125E-3 0.373309962441314
0.015625 0.36530592829283
0.03125 0.382986036503651
0.0625 0.383724970497049
0.125 0.36090290559056
0.25 0.343151183418341
};
\addlegendentry{$\varepsilon$-greedy}

\addplot[
color=blue,
solid,
mark=square,
thick,
]
table {
	0.03125 0.339103220903166
	0.0625 0.347138077607761
	0.125 0.34957005920592
	0.250 0.341277827182719
	0.500 0.340948157315732
	1.000 0.332140600260027
	2.000 0.3309382820282
};
\addlegendentry{UCB}

\end{axis}

\end{tikzpicture}
	\hspace*{20pt} \ref{legend_param_study}
	\caption{A parameter study of the $\varepsilon$-greedy and UCB algorithms. The average reward is obtained for the first $10000$ time steps with $k=500$ at $E_b/N_0=3.0$ dB.}
 	\label{fig:param}
\end{figure}

Fig.~\ref{fig:param} illustrates the dependence of the average reward on the parameters in $\varepsilon$-greedy and UCB algorithms for $\mathcal{P}(128,64)$. The simulation is carried out at $E_b/N_0 = 3.0$~dB and we set $k=500$ for all multi-armed bandit algorithms. In this figure, the average reward of the first $10000$ time steps received by the RL-CABP decoder is plotted against the parameter value. Note that a time step is increased by $1$ when the multi-armed bandit algorithm is required for the action selection, i.e., when CABP decoding has failed on the original factor-graph permutation $\pi_0$. As seen from Fig.~\ref{fig:param}, at $\varepsilon=2^{-4}$ and $c=2^{-3}$, RL-CABP decoding has the highest average reward value for $\varepsilon$-greedy and UCB algorithms, respectively. The TS algorithm does not require parameter tuning since $\alpha$ and $\beta$ parameters associated with each action are optimized during the decoding process.

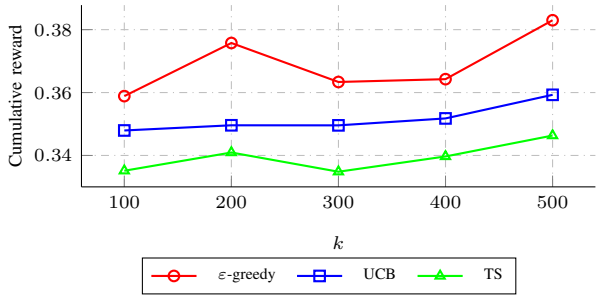
\begin{figure}[t!]
	\centering
	\begin{tikzpicture}
\pgfplotsset{
	label style = {font=\fontsize{7pt}{7}\selectfont},
	tick label style = {font=\fontsize{7pt}{7}\selectfont},
	log x ticks with fixed point/.style={
		xticklabel={
			\pgfkeys{/pgf/fpu=true}
			\pgfmathparse{exp(\tick)}%
			\pgfmathprintnumber[fixed relative, precision=3]{\pgfmathresult}
			\pgfkeys{/pgf/fpu=false}
		}
	},
	log y ticks with fixed point/.style={
		yticklabel={
			\pgfkeys{/pgf/fpu=true}
			\pgfmathparse{exp(\tick)}%
			\pgfmathprintnumber[fixed relative, precision=3]{\pgfmathresult}
			\pgfkeys{/pgf/fpu=false}
		}
	}
}

\begin{axis}[
log basis x=2,
yticklabel pos=left,
axis y line*=left,
ylabel style={yshift=-1em},
axis x line*=bottom,
ylabel=Cumulative reward,
xlabel=$k$,
ymajorgrids=true,
xmajorgrids=true,
ylabel style = {align=center},
grid style=dashdotted,
width=0.95\columnwidth, height=4cm,
legend style={
	at={(0,1e-5)},
	anchor=south west,
	column sep= 2mm,
	font=\fontsize{6pt}{7.2}\selectfont,
},
legend to name=legend_k,
legend columns=3,
],

\addplot[
color=red,
solid,
mark=o,
thick,
]
table{
	100 0.358867654765477	
	200 0.37575599879988	
	300 0.363359299329932
	400 0.364257576157616	
	500 0.382986036503651
};
\addlegendentry{$\varepsilon$-greedy}

\addplot[
color=blue,
solid,
mark=square,
thick,
]
table{
	100 0.34794281908191
	200 0.349590284128412	
	300 0.34957005920592
	400 0.351773792979299	
	500 0.35927999069907	
};
\addlegendentry{UCB}

\addplot[
color=green,
solid,
mark=triangle,
thick,
]
table{
	100 0.335138792179218	
	200 0.340933935293529	
	300 0.334828155615563	
	400 0.339682200820082
	500 0.346326458145814		
};
\addlegendentry{TS}

\end{axis}

\end{tikzpicture}
	\hspace*{20pt} \ref{legend_k}
	\caption{The impact of $k$ on the performance of different multi-armed bandit algorithms used by RL-CABP decoding for $\mathcal{P}(128,64)$, obtained for the first $10000$ time steps.}
	\label{fig:k}
	\vspace*{-1\baselineskip}
\end{figure}

Fig.~\ref{fig:k} illustrates the performance of multi-armed bandit algorithms used by RL-CABP decoding with different values of $k$. This simulation is also carried out at $E_b/N_0 = 3.0$~dB. We set $\varepsilon=2^{-4}$ for the $\varepsilon$-greedy algorithm and $c=2^{-3}$ for the UCB algorithm as those configurations provide the best performance in Fig.~\ref{fig:param}. It can be observed that for all the bandit algorithms, $k=500$ provides the largest cumulative reward after the first $10000$ time steps. Thus, we set $k=500$ for the rest of the paper.

\begin{figure}[t!]
	\centering
	\includegraphics[height=0.45\columnwidth,width=\columnwidth]{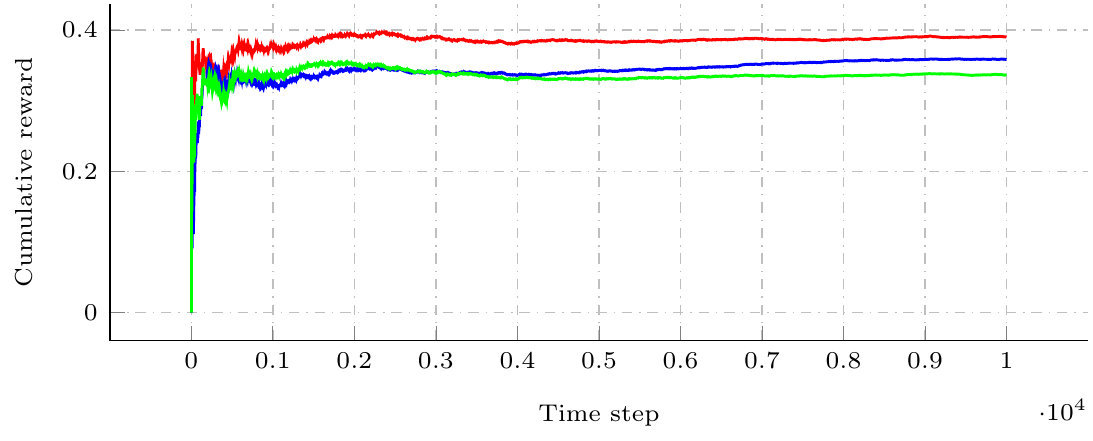}
	
	\hspace*{20pt} \includegraphics[height=0.55cm,width=4.5cm]{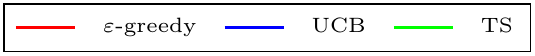}

	\caption{Performance comparison of various multi-armed bandit algorithms used by RL-CABP decoding. The simulation is obtained at $E_b/N_0=3.0$ dB with $k=500$, $\varepsilon=2^{-4}$, and $c=2^{-3}$.
	}
	\label{fig:reward}
	\vspace*{-1\baselineskip}
\end{figure}

Fig.~\ref{fig:reward} illustrates the average cumulative reward over the first $10000$ time steps for all the multi-armed bandit algorithms. The simulation is performed at $E_b/N_0=3.0$ dB with $k=500$, $\varepsilon=2^{-4}$, and $c=2^{-3}$. It can be seen that the $\varepsilon$-greedy algorithm has the best performance in terms of the average cumulative reward. In addition, the UCB algorithm performs slightly better than the TS algorithm. Note that the spikes in the early part of the curves are caused by the small value of the time step, which makes the calculation of the average cumulative reward unreliable at the initial phases of the algorithm.

Fig.~\ref{fig:fer:1} compares the FER of different factor-graph permutation selection schemes under the CABP decoding algorithm. In this figure, CABP denotes the CABP decoding algorithm performed only on the original factor-graph permutation. CP-CABP and RP-CABP denote the cyclically-shifted and random factor-graph permutations selection schemes proposed in \cite{hussami2009performance} and \cite{elkelesh2018belief}, respectively. Note that as there are $n=7$ cyclically-shifted permutations for $\mathcal{P}(128,64)$, we set the number of additional random permutations used by RP-CABP to $6$, and $M=7$ for the proposed RL-CABP decoder for a fair comparison. It can be seen that the proposed RL-CABP decoder under various multi-armed bandit algorithms has a similar FER performance. When compared with CP-CABP and RP-CABP, an error-correction performance gain of at least $0.125$~dB is obtained at the target FER of $10^{-4}$. In addition, an FER gain of around $0.62$~dB is obtained when the proposed RL-CABP decoding algorithm is compared with the baseline CABP decoder at the FER of $10^{-4}$.

\begin{figure}[t!]
	\centering
	\begin{tikzpicture}[spy using outlines = {rectangle, magnification=2.0, connect spies}]
\pgfplotsset{	
	label style = {font=\fontsize{7pt}{7}\selectfont},
	tick label style = {font=\fontsize{7pt}{7}\selectfont}
}

\begin{axis}[
scale = 1,
ymode=log,
xlabel={$E_b/N_0$ [\text{dB}]}, xlabel style={yshift=0.8em},
ylabel={FER}, ylabel style={yshift=-0.75em},
xtick={4.5,4.75,5,5.25,5.5,5.75,6,6.25,6.5},
grid=both,
ymajorgrids=true,
xmajorgrids=true,
grid style=dashdotted,
width=1\columnwidth, height=5.25cm,
thick,
mark size=2,
legend cell align={left},
legend style={
	at={(0,1e-5)},
	anchor=south west,
	column sep= 2mm,
	font=\fontsize{6pt}{7.2}\selectfont,
},
  legend to name=legend_fer_comp2,
  legend columns=3,
]

\addplot[
color=black,
solid,
mark=*,
mark options={solid},
thick,
mark size=2,
]
table {
	4.5	0.01155012
	5	0.00345449
	5.5	0.00078718
	6	0.0001558
	6.5	3.2E-05
};
\addlegendentry{BP \cite{yuan2014early}}

\addplot[
color=gray,
solid,
mark=square*,
mark options={solid},
thick,
mark size=2,
]
table {
	4.5	0.00226002
	5	0.00039276
	5.5	5.139E-05
};
\addlegendentry{CA-SCL2 \cite{Alexios_LLR_SCLD}}

\addplot[
color=cyan,
solid,
mark=o,
mark options={solid},
thick,
mark size=2,
]
table {
	4.5	0.00038522
	5	3.861E-05
};
\addlegendentry{CA-SCL4 \cite{Alexios_LLR_SCLD}}

\addplot[
color=green,
solid,
mark=diamond*,
mark options={solid},
thick,
mark size=2.5,
]
table {
	4.5	0.00145575
	5	0.00023423
	5.5	3.5E-05
};
\addlegendentry{RL-CABP (TS)}

\addplot[
color=blue,
solid,
mark=triangle*,
mark options={solid},
thick,
mark size=2.5,
]
table {
	4.5	0.0012864
	5	0.00022933
	5.5	3.34E-05
};
\addlegendentry{RL-CABP (UCB)}

\addplot[
color=red,
solid,
mark=star,
mark options={solid},
thick,
mark size=2.5,
]
table {
	4.5	0.00124935
	5	0.00021724
	5.5	3.2E-05
};
\addlegendentry{RL-CABP ($\varepsilon$-greedy)}

\end{axis}


\end{tikzpicture}
	\hspace*{10pt} \ref{legend_fer_comp2}
	\caption{Error-correction performance of RL-CABP decoding and other decoding algorithms of polar codes.}
	\label{fig:fer:2}
	\vspace*{-1\baselineskip}
\end{figure}
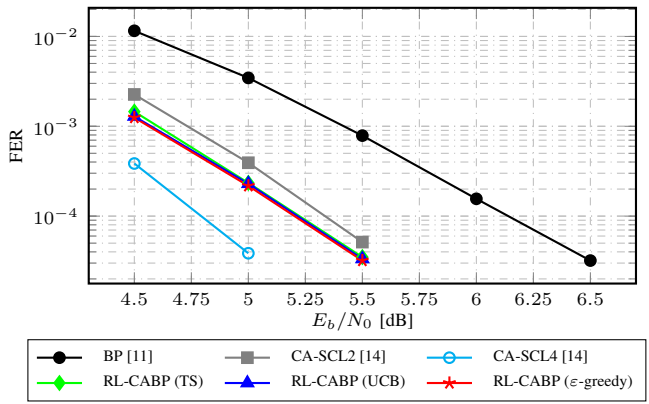

Fig.~\ref{fig:fer:2} compares the error-correction performance of the proposed RL-CABP decoding with BP decoding and CA-SCL decoding of polar codes. In Fig.~\ref{fig:fer:2}, CA-SCL$L$ indicates the CA-SCL decoder with a list size of $L$. It can be observed that at the target FER of $10^{-4}$, the FER performance of the proposed RL-CABP decoder is around $0.92$~dB better than that of the BP decoding algorithm in \cite{yuan2014early}. At the same target FER, CA-SCL$4$ provides a better error-correction performance in comparison with the proposed RL-CABP decoder. However, compared with CA-SCL$2$ at the same target FER, the proposed decoder has a performance gain of around $0.12$~dB, under different multi-armed bandit algorithms.

\begin{figure}[t!]
	\centering
	\begin{tikzpicture}[spy using outlines = {rectangle, magnification=2.0, connect spies}]
\pgfplotsset{	
	label style = {font=\fontsize{7pt}{7}\selectfont},
	tick label style = {font=\fontsize{7pt}{7}\selectfont}
}

\begin{axis}[
scale = 1,
ymode=log,
xlabel={$E_b/N_0$ [\text{dB}]}, xlabel style={yshift=0.8em},
ylabel={FER}, ylabel style={yshift=-0.75em},
xtick={4.5,4.75,5,5.25,5.5,5.75,6},
grid=both,
ymajorgrids=true,
xmajorgrids=true,
grid style=dashdotted,
width=1\columnwidth, height=5.25cm,
thick,
mark size=2,
legend cell align={left},
legend style={
	at={(0,1e-5)},
	anchor=south west,
	column sep= 2mm,
	font=\fontsize{6pt}{7.2}\selectfont,
},
  legend to name=legend_fer_comp1,
  legend columns=3,
]

\addplot[
color=black,
solid,
mark=*,
mark options={solid},
thick,
mark size=2,
]
table {
	4.5	0.006645
	5	0.00161722
	5.5	0.00031129
	6	5.48E-05
};
\addlegendentry{CABP \cite{Doan_ICC19}}

\addplot[
color=gray,
solid,
mark=square*,
mark options={solid},
thick,
mark size=2,
]
table {
	4.5	0.0020539
	5	0.00040173
	5.5	5.72E-05
};
\addlegendentry{CP-CABP \cite{hussami2009performance}}

\addplot[
color=cyan,
solid,
mark=o,
mark options={solid},
thick,
mark size=2,
]
table {
	4.5	0.00187101
	5	0.00033595
	5.5	5.15E-05
};
\addlegendentry{RP-CABP \cite{elkelesh2018belief}}

\addplot[
color=green,
solid,
mark=diamond*,
mark options={solid},
thick,
mark size=2.5,
]
table {
	4.5	0.00145575
	5	0.00023423
	5.5	3.5E-05
};
\addlegendentry{RL-CABP (TS)}

\addplot[
color=blue,
solid,
mark=triangle*,
mark options={solid},
thick,
mark size=2.5,
]
table {
	4.5	0.0012864
	5	0.00022933
	5.5	3.34E-05
};
\addlegendentry{RL-CABP (UCB)}

\addplot[
color=red,
solid,
mark=star,
mark options={solid},
thick,
mark size=2.5,
]
table {
	4.5	0.00124935
	5	0.00021724
	5.5	3.2E-05
};
\addlegendentry{RL-CABP ($\varepsilon$-greedy)}

\end{axis}

\end{tikzpicture}
	\hspace*{10pt} \ref{legend_fer_comp1}
	\caption{Error-correction performance of different factor-graph permutation selection schemes for $\mathcal{P}(128,64)$.}
	\label{fig:fer:1}
	\vspace*{-0.5\baselineskip}
\end{figure}
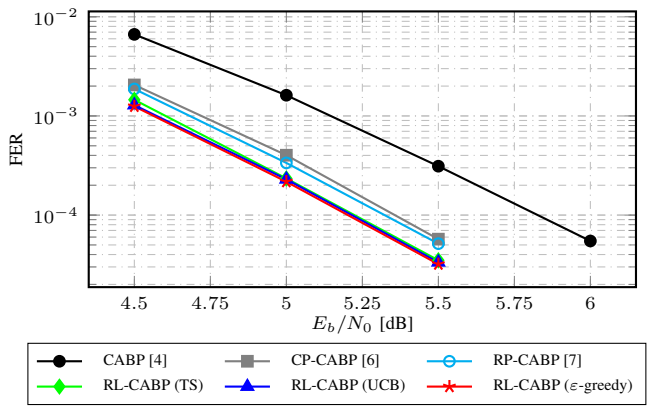

Table~\ref{tab:complx} shows the maximum number of computations required by various permutation selection schemes used in Fig.~\ref{fig:fer:1}. Among all the multi-armed bandit algorithms, the $\varepsilon$-greedy algorithm in general has the lowest computational complexity. This is because the TS algorithm requires a sampling process for $k$ different $\Beta$ distributions, which in general requires higher computational complexity than applying an i.i.d. sampling from the interval of $(0,1)$ and doing a multiplication as required by the $\varepsilon$-greedy algorithm. In addition, although using the cyclically-shifted factor-graph permutations does not consume any additional complexity for the factor-graph permutation selection, this technique is not applicable when more than $n$ different permutations are required. It can also be observed that the main drawback of the multi-armed bandit algorithms is the sorting operations required to identify the exploitation action. However, as described in Section~\ref{sec:RL-CABPB}, the action selection process can be performed in parallel with the first CABP decoding attempt. Therefore, there is no additional latency overhead. Furthermore, the approaches in \cite{hussami2009performance} and \cite{elkelesh2018belief} come with the cost of error-correction performance degradation when compared with the proposed RL-CABP decoder as illustrated in Fig.~\ref{fig:fer:1}.

\begin{table}[t!]
	\caption{Computational complexity of different permutation selection schemes in terms of the maximum number of operations performed}
	\centering
	\begin{tabular}{l c c c c c}
		\toprule
		Operations &  \cite{hussami2009performance} & \cite{elkelesh2018belief} & $\varepsilon$-greedy & UCB & TS \\
		\midrule
		$+$ & 0 & 0 & 2 & 2 + $k$& 2 \\
		$-$ & 0 & 0 & 1 & 1 & 0 \\
		$\times$ & 0 & 0 & 1 & 1+$k$ & 0 \\
		$\divisionsymbol$ & 0 & 0 & 0 & $k$ & 0 \\		
		$\sqrt{\color{white}{.}}$ & 0 & 0 & 0 & $k$ & 0 \\
		$\ln$ & 0 & 0 & 0 & $k$ & 0 \\
		Random sampling & 0 & $M-1$ & 1 & 0 & $k$ \\
		Sorting & 0 & 0 & $k$ & $k$ & $k$\\
		\bottomrule
	\end{tabular}
	\label{tab:complx}
	\vspace*{-1\baselineskip}
\end{table}


\section{Conclusions}
\label{sec:conclude}
In this paper, we first showed that the selection of factor-graph permutations for polar decoding can be formalized as a multi-armed bandit problem in RL. We then proposed an RL-CABP decoding algorithm that utilizes the state-of-the-art algorithms for the multi-armed bandit problem to select the factor-graph permutations under CABP decoding of polar codes. We showed that for a 5G polar code of length $128$, with $64$ information bits and concatenated with a $16$-bit 5G CRC, the FER of the proposed decoder is around $0.125$~dB better than that of the technique that selects the factor-graph permutations randomly, at the target FER of $10^{-4}$. In addition, we showed that there is no additional latency overhead for the selection of factor-graph permutations of the proposed decoder compared with the approach that selects the factor-graph permutations at random.

\section*{Acknowledgment}

S. A. Hashemi is supported by a Postdoctoral Fellowship from the Natural Sciences and Engineering Research Council of Canada (NSERC).


\begin{thebibliography}{10}
	\providecommand{\url}[1]{#1}
	\csname url@samestyle\endcsname
	\providecommand{\newblock}{\relax}
	\providecommand{\bibinfo}[2]{#2}
	\providecommand{\BIBentrySTDinterwordspacing}{\spaceskip=0pt\relax}
	\providecommand{\BIBentryALTinterwordstretchfactor}{4}
	\providecommand{\BIBentryALTinterwordspacing}{\spaceskip=\fontdimen2\font plus
		\BIBentryALTinterwordstretchfactor\fontdimen3\font minus
		\fontdimen4\font\relax}
	\providecommand{\BIBforeignlanguage}[2]{{%
			\expandafter\ifx\csname l@#1\endcsname\relax
			\typeout{** WARNING: IEEEtran.bst: No hyphenation pattern has been}%
			\typeout{** loaded for the language `#1'. Using the pattern for}%
			\typeout{** the default language instead.}%
			\else
			\language=\csname l@#1\endcsname
			\fi
			#2}}
	\providecommand{\BIBdecl}{\relax}
	\BIBdecl
	
	\bibitem{arikan}
	E.~Ar{\i}kan, ``Channel polarization: A method for constructing
	capacity-achieving codes for symmetric binary-input memoryless channels,''
	\emph{{IEEE} Trans. Inf. Theory}, vol.~55, no.~7, pp. 3051--3073, July 2009.
	
	\bibitem{tal_list}
	I.~Tal and A.~Vardy, ``List decoding of polar codes,'' \emph{{IEEE} Trans. Inf.
		Theory}, vol.~61, no.~5, pp. 2213--2226, March 2015.
	
	\bibitem{3gpp_report}
	\BIBentryALTinterwordspacing
	3GPP, ``Multiplexing and channel coding ({R}elease 10) {3GPP} {TS} {21.101}
	{v10.4.0.}'' Oct. 2018. [Online]. Available:
	\url{http://www.3gpp.org/ftp/Specs/2018-09/Rel-10/21\_series/21101-a40.zip}
	\BIBentrySTDinterwordspacing
	
	\bibitem{Doan_ICC19}
	N.~{Doan}, S.~A. {Hashemi}, E.~N. {Mambou}, T.~{Tonnellier}, and W.~J. {Gross},
	``Neural belief propagation decoding of {CRC}-polar concatenated codes,''
	\emph{IEEE Int. Conf. on Commun.}, pp. 1--6, May 2019.
	
	\bibitem{CABPList}
	\BIBentryALTinterwordspacing
	M.~Geiselhart, A.~Elkelesh, M.~Ebada, S.~Cammerer, and S.~ten Brink,
	``{CRC}-aided belief propagation list decoding of polar codes,'' \emph{IEEE
		Int. Sym. on Inf. Theory (to appear).}, 2020. [Online]. Available:
	\url{https://arxiv.org/abs/2001.05303}
	\BIBentrySTDinterwordspacing
	
	\bibitem{hussami2009performance}
	N.~Hussami, S.~B. Korada, and R.~Urbanke, ``Performance of polar codes for
	channel and source coding,'' in \emph{IEEE Int. Symp. on Inf. Theory}, 2009,
	pp. 1488--1492.
	
	\bibitem{elkelesh2018belief}
	A.~Elkelesh, M.~Ebada, S.~Cammerer, and S.~ten Brink, ``Belief propagation
	decoding of polar codes on permuted factor graphs,'' in \emph{IEEE Wireless
		Commun. and Net. Conf.}, April 2018, pp. 1--6.
	
	\bibitem{Doan_GLOBECOM}
	N.~{Doan}, S.~A. {Hashemi}, M.~{Mondelli}, and W.~J. {Gross}, ``On the decoding
	of polar codes on permuted factor graphs,'' \emph{IEEE Global Commun. Conf.},
	pp. 1--6, Dec 2018.
	
	\bibitem{LoopSimp}
	Y.~{Ren}, Y.~{Shen}, Z.~{Zhang}, X.~{You}, and C.~{Zhang}, ``Efficient belief
	propagation polar decoder with loop simplification based factor graphs,''
	\emph{IEEE Transactions on Vehicular Technology}, pp. 1--1, 2020.
	
	\bibitem{arikan2010polar}
	E.~Ar{\i}kan, ``Polar codes: A pipelined implementation,'' in \emph{Proc. 4th
		Int. Symp. on Broad. Commun.}, 2010, pp. 11--14.
	
	\bibitem{yuan2014early}
	B.~Yuan and K.~K. Parhi, ``Early stopping criteria for energy-efficient
	low-latency belief-propagation polar code decoders,'' \emph{IEEE Transactions
		on Signal Processing}, vol.~62, no.~24, pp. 6496--6506, Dec. 2014.
	
	\bibitem{Sutton}
	R.~S. Sutton and A.~G. Barto, \emph{Reinforcement Learning: An
		Introduction}.\hskip 1em plus 0.5em minus 0.4em\relax Cambridge, MA, USA: A
	Bradford Book, 2018.
	
	\bibitem{TS}
	S.~Agrawal and N.~Goyal, ``Analysis of thompson sampling for the multi-armed
	bandit problem,'' in \emph{Conf. on Learning Theory}, 2012, pp. 39--1.
	
	\bibitem{Alexios_LLR_SCLD}
	A.~Balatsoukas-Stimming, M.~B. Parizi, and A.~Burg, ``{LLR}-based successive
	cancellation list decoding of polar codes,'' \emph{{IEEE} Trans. Signal
		Process.}, vol.~63, no.~19, pp. 5165--5179, Oct. 2015.
	
\end{thebibliography}

\end{document}